\documentclass[12pt]{article}
\usepackage{amsmath}
\usepackage{amssymb}

\def \be{\begin{equation}}
\def \ee{\end{equation}}
\def \beq{\begin{eqnarray}}
\def \eeq{\end{eqnarray}}
\begin{document}

\date{\today }
\date{\today }

\bigskip

\bigskip \ 

\bigskip \ 

\begin{center}
\textbf{Higher dimensional Charged Black Holes }

\ 

\textbf{as constrained systems}\bigskip \ 

\smallskip

J. A. Nieto$^{\dag }$\footnote{%
niet@uas.edu.mx, janieto1@asu.edu}, E. A. Leon$^{\dag }$\footnote{%
ealeon@uas.edu.mx}, V. M. Villanueva$^{\ddag }$\footnote{%
vvillanu@ifm.umich.mx}

\smallskip

$^{\dag}$\textit{Facultad de Ciencias F\'{\i}sico-Matem\'{a}ticas de la
Universidad Aut\'{o}noma}

\textit{de Sinaloa, 80010, Culiac\'{a}n Sinaloa, M\'{e}xico}

\smallskip

$^{\ddag }$\textit{Instituto de F\'{\i}sica y Matem\'{a}ticas de la
Universidad Michoacana de San}

\textit{\ Nicol\'{a}s de Hidalgo, Morelia Michoac\'{a}n, M\'{e}xico}

\bigskip \ 

\bigskip \ 

\textbf{\bigskip \ }

\textbf{Abstract}
\end{center}

We construct a Lagrangian and Hamiltonian formulation for charged black
holes in a d-dimensional maximally symmetric spherical space. By considering
first new variables that give raise to an interesting dimensional reduction
of the problem, we show that the introduction of a charge term is compatible
with classical solutions to Einstein equations. In fact, we derive the
well-known solutions for charged black holes, specially in the case of d=4,
where the Reissner-Nordstr\"{o}m solution holds, without reference to
Einstein field equations. We argue that our procedure may be of help for
clarifying symmetries and dynamics of black holes, as well as some quantum
aspects.

\bigskip \bigskip

\ 

\bigskip \  \ 


Pacs numbers: 04.20.Jb, 04.50.-Gh, 11.10.-Kk, 11.30.Ly

February 2013

\newpage

\noindent \textbf{1. Introduction}\smallskip

\noindent The static solutions of Einstein field equations associated with
spherical symmetric spaces have been extensively studied in the literature.
However, in the past decades, mainly due to the emergence of string theory,
there has been a renewed interest in studying black hole solutions involving
extra dimensions [1]-[7]. In fact, the study of black holes in several
dimensions is a strong indicative of the level of consistency developed in
different approaches for solving the problem of quantum gravity [8]-[12].

In this work, we focus our attention in a method for obtaining the static
solutions for charged black holes, that permit us to make the analysis in
complete analogy to that of a relativistic point particle. In particular, we
find that charged black holes can be treated as a constrainned hamiltonian
system.

It is well known that, by taking the radial coordinate as an evolution
parameter, one can describe the black hole dynamics in terms of a type of
Hamiltonian. Recently, however, such Hamiltonian structure was described in
such a way that it establishes a clear connection with constrained
Hamiltonian systems (see \cite{Nieto1} and references therein). In this work
we start with a very general form for the metric in order to generalize such
constrained Hamiltonian system to higher dimensions.

Moreover, starting with the Einstein-Maxwell action, we prove that our
approach can be extended to the case of charged black holes. In this case we
verify that one can recover the Reissner-Nordstr\"{o}m solution. One of the
advantages of our procedure is that one can perform a similar analysis to
the relativistic point particle. For instance, the invariance of the
corresponding action under an arbitrary change of the radial parameter can
be linked to a Lagrange multiplier. In this way, fixing this Lagrange
multiplier is equivalent to set a gauge in order to recover the
Reissner-Nordstr\"{o}m solution. Moreover, considering new variables the
Lagrangian formulation is related to a Hamiltonian formulation that allows
to treat the problem of charged black holes as a constrained system. We
argue that this procedure may help to tackle the problem of quantum black
holes \cite{Govaerts}\cite{Henneaux}.

The structure of this paper is the following: In section 2, we briefly
review the general method of obtaining the Reissner-Nordstr\"{o}m solution
from the Maxwell-Einstein field equations. In section 3, we introduce the
Lagrangian formulation in terms of a particular set of variables that
simplify the analysis. All the developments made until this point lead us,
in section 4, to introduce an associated Hamiltonian which allows to take a
view of the charged black holes in higher dimensions as a constrained
Hamiltonian system. We conclude with various remarks, in section 5,
summarizing the work and mentioning some perspectives for future
explorations on the subject. In particular, we remark that our analysis can
be useful for quantum black holes in higher dimensions.

\smallskip

\noindent \textbf{2. The Einstein-Maxwell model for charged black holes}

\smallskip

\noindent We start by considering the Einstein-Maxwell action in $d$
dimensions, given by \cite{KunzNavarro}

\begin{equation}
S=\frac{1}{16\pi G_{d}}\int_{M^{d}}\sqrt{-\gamma }\left( R-F_{\mu \nu
}F^{\mu \nu }\right) ,  \tag{1}
\end{equation}%
where $G_{d}$ is the gravitational constant in $d$ dimensions, $R$ is the
scalar curvature. Moreover, $F_{\mu \nu }=\partial _{\mu }A_{\nu }-\partial
_{\nu }A_{\mu }$ is the electromagnetic field strength tensor defined in
terms of a potential $A_{\mu }$, and $\gamma $ is the determinant of the
metric $\gamma _{\mu \nu }$. To guarantee static spherical symmetry, we
assume the general metric

\begin{equation}
ds^{2}=-e^{f(r)}dt^{2}+e^{h(r)}dr^{2}+\varphi ^{2}(r)\tilde{\gamma}_{ij}(\xi
^{k})d\xi ^{i}d\xi ^{j}.  \tag{2}
\end{equation}%
Here, we have specified the functional form for $f$, $h$ and $\varphi $ in
terms of $r$ and considered the speed of light $c=1$. The submetric $\tilde{%
\gamma}_{ij}(\xi ^{k})$ corresponds to a maximally symmetric subspace in $%
(d-2)$-dimensions with curvilinear coordinates $\xi ^{i}$ that are
independent of time $t$ and $r$. The notation in this article is as follows:
greek indices like $\mu ,\nu $ run from $0$ to $d-1$, while latin indices
like $i,j$ run from $2$ to $d-1$; also, as a convenient notation we will be
referring to derivatives respect to $r$ with dot notation, that is $\dot{f}%
\equiv \frac{df}{dr}$ and $\ddot{f}\equiv \frac{d^{2}f}{dr^{2}}$, for
instance.

Here, we shall consider electrically charged black holes, with the only
non-vanishing strength field tensor component $F_{10}=-F_{01}=\partial
_{1}A_{0}=\dot{\chi}$, where $\chi $ is the electric potential.

Taking into account the form of the metric (2), we have the relation between
determinants $\sqrt{-\gamma }=e^{\frac{f+h}{2}}\varphi ^{(d-2)}\sqrt{\tilde{%
\gamma}}$ and also $F_{\lambda \tau }F^{\lambda \tau }=-2e^{-(h+f)}\dot{\chi}%
^{2}$. In this manner, by using the curvature scalar $R$ corresponding to
the metric (2) (cf. Appendix), we find that the action (1), up to a total
derivative, takes the form%
\begin{equation}
\begin{array}{c}
S=\frac{(d-2)}{16\pi G_{d}}\int_{M^{d}}\sqrt{\tilde{\gamma}}\big \{ \varphi
^{(d-2)}e^{\frac{f-h}{2}}\left[ (d-3)\frac{\dot{\varphi}^{2}}{\varphi ^{2}}+%
\dot{f}\frac{\dot{\varphi}}{\varphi }\right] \\ 
\\ 
+k(d-3)e^{\frac{f+h}{2}}\varphi ^{(d-4)}+\frac{2}{(d-2)}e^{\frac{-(f+h)}{2}%
}\varphi ^{(d-2)}\dot{\chi}^{2}\big \}.%
\end{array}
\tag{3}
\end{equation}

We shall focus our attention on the case $d\geq 4$. If we further define $%
\mathcal{F}\equiv e^{\frac{f}{2}}$ and $\Omega \equiv e^{\frac{h}{2}}$, then
we can express (3) as%
\begin{equation}
\begin{array}{c}
S=\frac{(d-2)}{16\pi G_{d}}\int_{M^{d}}\sqrt{\tilde{\gamma}}\big \{ \mathcal{%
F}\Omega ^{-1}\varphi ^{(d-4)}\left[ (d-3)\dot{\varphi}^{2}+2\frac{\mathcal{%
\dot{F}}}{\mathcal{F}}\dot{\varphi}\varphi \right] \\ 
\\ 
+k(d-3)\mathcal{F}\Omega \varphi ^{(d-4)}+\frac{2}{(d-2)}\mathcal{F}%
^{-1}\Omega ^{-1}\varphi ^{(d-2)}\dot{\chi}^{2}\big \}.%
\end{array}
\tag{4}
\end{equation}%
The variation of $S$ respect to $\mathcal{F}$, $\Omega $ and $\varphi $
yields%
\begin{equation}
2\frac{\ddot{\varphi}}{\varphi }-2\frac{\dot{\varphi}\dot{\Omega}}{\varphi
\Omega }+(d-3)\frac{\dot{\varphi}^{2}}{\varphi ^{2}}-k(d-3)\Omega
^{2}\varphi ^{-2}+\frac{2}{(d-2)}\mathcal{F}^{-2}\dot{\chi}^{2}=0,  \tag{5}
\end{equation}

\begin{equation}
(d-3)\frac{\dot{\varphi}^{2}}{\varphi ^{2}}+2\frac{\mathcal{\dot{F}}}{%
\mathcal{F}}\frac{\dot{\varphi}}{\varphi }-k(d-3)\Omega ^{2}\varphi ^{-2}+%
\frac{2}{(d-2)}\mathcal{F}^{-2}\dot{\chi}^{2}=0  \tag{6}
\end{equation}%
and%
\begin{equation}
\frac{(d-4)(d-3)}{2}\left( \frac{\dot{\varphi}^{2}}{\varphi ^{2}}-k\Omega
^{2}\varphi ^{-2}\right) +(d-3)\left[ \frac{\dot{\varphi}}{\varphi }\left( 
\frac{\mathcal{\dot{F}}}{\mathcal{F}}-\frac{\dot{\Omega}}{\Omega }\right) +%
\frac{\ddot{\varphi}}{\varphi }\right] -\frac{\mathcal{\dot{F}}\dot{\Omega}}{%
\mathcal{F}\Omega }+\frac{\mathcal{\ddot{F}}}{\mathcal{F}}-\mathcal{F}^{-2}%
\dot{\chi}^{2}=0,  \tag{7}
\end{equation}%
respectively, while the variation respect to $\chi $ gives%
\begin{equation}
\frac{d}{dr}\{ \mathcal{F}^{-1}\Omega ^{-1}\varphi ^{(d-2)}\dot{\chi}\}=0. 
\tag{8}
\end{equation}

Now, by combining (6) and (5) we get the relation%
\begin{equation}
\frac{d}{dr}\left \{ \ln (\mathcal{F}^{-1}\Omega ^{-1}\dot{\varphi})\right
\} =0,  \tag{9}
\end{equation}%
We observe that (8) and (9) imply%
\begin{equation}
\mathcal{F}^{-1}\Omega ^{-1}\varphi ^{(d-2)}\dot{\chi}=A  \tag{10}
\end{equation}%
and

\begin{equation}
\mathcal{F}^{-1}\Omega ^{-1}\dot{\varphi}=B,  \tag{11}
\end{equation}%
respectively, where $A$ and $B$ are integration constants. Thus, from (10)
and (11) we obtain a relation between the electric radial field $\dot{\chi}$
and the function $\varphi $, namely%
\begin{equation}
\dot{\chi}=C\frac{\dot{\varphi}}{\varphi ^{(d-2)}},  \tag{12}
\end{equation}%
where $C=\frac{A}{B}$.

Of course, these results can be obtained if one starts with Einstein-Maxwell
field equations. In fact, the Einstein field equations can be obtained by
making variations of the action (1) respect to the metric $\gamma $,
resulting%
\begin{equation}
G_{\mu \nu }=R_{\mu \nu }-\frac{1}{2}\gamma _{\mu \nu }R=2T_{\mu \nu }, 
\tag{13}
\end{equation}%
where the energy momentum tensor $T_{\mu \nu }$ is%
\begin{equation}
T_{\mu \nu }=F_{\mu \lambda }F_{\nu }^{~\lambda }-\frac{1}{4}\gamma _{\mu
\nu }F_{\lambda \tau }F^{\lambda \tau }.  \tag{14}
\end{equation}

Considering the metric (2) we note that the first two diagonal elements of
the energy-momentum tensor are given by $T_{00}=\frac{1}{2}\Omega ^{-2}\dot{%
\chi}^{2}$ and $T_{11}=-\frac{1}{2}\mathcal{F}^{-2}\dot{\chi}^{2}$. Taking
into account these expressions for $T_{00}$ and $T_{11}$, and adding
components $G_{00}$ and $G_{11}$ of the field equations in (13), we see that
the following relation holds:%
\begin{equation}
\Omega ^{2}R_{00}+\mathcal{F}^{2}R_{11}=0.  \tag{15}
\end{equation}%
If we substitute $R_{00}$ and $R_{11}$ from the Appendix, we obtain again
(9).

By making variations of (1) with respect the gauge field $A_{\mu }$ the
Maxwell field equation can also be obtained;%
\begin{equation}
\nabla _{\mu }F^{\mu \nu }=0,  \tag{16}
\end{equation}%
where $\nabla _{\mu }$ stands for covariant derivative. By taking the zero
component of the Maxwell equation (16) we obtain basically (8) after a
direct computation, and this in turn implies the solution (10).

\smallskip

\noindent \textbf{3. Charged black holes and Lagrangian formalism}

\smallskip

\noindent Variational methods for the case of black holes have been used
extensively in the past [15]-[18]. Now we present a direct method that takes
the radial parameter $r$ as the fundamental independent variable. Getting
back to the initial action, we remark that the integrand in (4) can be
interpreted as a type of Lagragian $\mathcal{L}$. We introduce the
`coordinates' $q^{a}\in \{q^{1},q^{2},q^{3}\}$ given by $\varphi =e^{q^{1}}$%
, $\mathcal{F}=e^{q^{2}}$, $\chi =q^{3}$, in such a way that $\mathcal{L}$
can be written as

\begin{equation}
\mathcal{L}=\frac{1}{2}\lambda ^{-1}\left[ (d-3)\left( \dot{q}^{1}\right)
^{2}+2\dot{q}^{1}\dot{q}^{2}+\frac{2}{(d-2)}e^{-2q^{2}}\left( \dot{q}%
^{3}\right) ^{2}\right] +\frac{1}{2}\lambda m_{o}^{2},  \tag{17}
\end{equation}%
where%
\begin{equation}
\lambda =\varphi ^{-(d-2)}\mathcal{F}^{-1}\Omega ,  \tag{18}
\end{equation}%
which plays the role of a Lagrange multiplier, and 
\begin{equation}
m_{o}^{2}=k(d-3)\mathcal{F}^{2}\varphi ^{2(d-3)}  \tag{19}
\end{equation}%
is the analogue of a mass term.

The Euler-Lagrange equations obtained from (17), by making variations of $%
q^{1},q^{2}$ and $q^{3}$ respectively, become%
\begin{equation}
\frac{d}{dr}\left \{ \lambda ^{-1}\left[ (d-3)\dot{q}^{1}+\dot{q}^{2}\right]
\right \} -\lambda (d-3)m_{o}^{2}=0,  \tag{20}
\end{equation}%
\begin{equation}
\frac{d}{dr}\left \{ \lambda ^{-1}\dot{q}^{1}\right \} +\lambda ^{-1}\frac{2%
}{(d-2)}e^{-2q^{2}}\left( \dot{q}^{3}\right) ^{2}-\lambda m_{o}^{2}=0, 
\tag{21}
\end{equation}%
and%
\begin{equation}
\frac{d}{dr}\left[ \lambda ^{-1}e^{-2q^{2}}\left( \dot{q}^{3}\right) \right]
=0.  \tag{22}
\end{equation}%
The variation respect to $\lambda $ gives%
\begin{equation}
\lambda ^{-2}\left[ (d-3)\left( \dot{q}^{1}\right) ^{2}+2\dot{q}^{1}\dot{q}%
^{2}+\frac{2}{(d-2)}e^{-2q^{2}}\left( \dot{q}^{3}\right) ^{2}\right]
-m_{o}^{2}=0.  \tag{23}
\end{equation}

Before we continue, it is worthwhile mentioning that the equations (20)-(23)
are consistent with equations (5)-(8). In fact, using the definition of the
coordinates $\left \{ q^{a}\right \} $ and performing the corresponding
derivatives, we note first that (22) is just (8), and also that, up to a
multiplicative factor, (23) is equivalent to (6). Now, multiplying (21) by $%
(d-3)$ and substracting the result to (20), we obtain%
\begin{equation}
\frac{d}{dr}\left \{ \lambda ^{-1}\dot{q}^{2}\right \} -2\lambda ^{-1}\left( 
\frac{d-3}{d-2}\right) e^{-2q^{2}}\left( \dot{q}^{3}\right) ^{2}=0.  \tag{24}
\end{equation}%
After some computation it can be shown that (7) follows from (21), (23) and\
(24). Finally, it is straightforward to get the equation (5) from (21) and
(23).

Equation (22) can be solved and gives%
\begin{equation}
\lambda ^{-1}e^{-2q^{2}}\dot{q}^{3}=A,  \tag{25}
\end{equation}%
which is equivalent to (10). Using (24) and (25) one gets%
\begin{equation}
\frac{d}{dr}\left \{ \lambda ^{-1}\dot{q}^{2}\right \} -2A\left( \frac{d-3}{%
d-2}\right) \dot{q}^{3}=0,  \tag{26}
\end{equation}%
This relation will be useful below. Now, equations (20) and (23) imply that%
\begin{equation}
\frac{d}{dr}\left \{ \lambda ^{-1}\left[ (d-3)\dot{q}^{1}+\dot{q}^{2}\right]
\right \} =(d-3)\lambda ^{-1}\left[ (d-3)\left( \dot{q}^{1}\right) ^{2}+2%
\dot{q}^{1}\dot{q}^{2}+\frac{2}{(d-2)}e^{-2q^{2}}\left( \dot{q}^{3}\right)
^{2}\right] .  \tag{27}
\end{equation}%
Making use of (24) in (27) we get, after simplifications,%
\begin{equation}
\frac{d}{dr}\left[ \ln \left( \lambda ^{-1}\dot{q}^{1}\right)
-(d-3)q^{1}-2q^{2}\right] =0.  \tag{28}
\end{equation}%
Considering the definitions of $(q^{1},q^{2},q^{3})$ in terms of $(\varphi ,%
\mathcal{F},\chi )$, equations (26) and (28) leads to%
\begin{equation}
\lambda ^{-1}\frac{\mathcal{\dot{F}}}{\mathcal{F}}-2A\left( \frac{d-3}{d-2}%
\right) \chi =D  \tag{29}
\end{equation}%
and 
\begin{equation}
\lambda ^{-1}\dot{\varphi}\varphi ^{-(d-2)}\mathcal{F}^{-2}=B,  \tag{30}
\end{equation}%
respectively. Here, $D$ and $B$ are integration constants in agreement with
(11). Relations (25) and (30) provide a solution for the electric field $%
\dot{\chi}$ in terms of $\dot{\varphi}$ and $\varphi $: 
\begin{equation}
\dot{\chi}=\frac{A}{B}\frac{\dot{\varphi}}{\varphi ^{(d-2)}},  \tag{31}
\end{equation}%
which is just (12). By combining (30) with this equation we obtain that $%
\lambda ^{-1}=A\frac{\mathcal{F}^{2}}{\dot{\chi}}$, which can be inserted
into (29), resulting in%
\begin{equation}
A\mathcal{F\dot{F}}-2A\left( \frac{d-3}{d-2}\right) \chi \dot{\chi}=D\dot{%
\chi}.  \tag{32}
\end{equation}%
One can express (32) as

\begin{equation}
\frac{d}{dr}[A\mathcal{F}^{2}-2D\chi -2A\left( \frac{d-3}{d-2}\right) \chi
^{2}]=0.  \tag{33}
\end{equation}%
Now, from (31) one notes that

\begin{equation}
\chi =\frac{-C}{(d-3)\varphi ^{(d-3)}}+G,  \tag{34}
\end{equation}%
where $G$ is integration constant. Thus, by substituting (34) into (33) one
obtains

\begin{equation}
\begin{array}{c}
\mathcal{F}^{2}=\left[ \frac{E}{A}+\frac{2DG}{A}+2\left( \frac{d-3}{d-2}%
\right) G^{2}\right] \\ 
\\ 
-\left( \frac{2D}{B(d-3)}+\frac{4CG}{(d-2)}\right) \frac{1}{\varphi ^{(d-3)}}%
+\frac{2C^{2}}{(d-2)(d-3)\varphi ^{2(d-3)}}.%
\end{array}
\tag{35}
\end{equation}%
Here, the quantity $E$ is another integration constant.

Recalling that $\mathcal{F}=e^{\frac{f}{2}}$, then the $\gamma _{00}$
component of the metric $\gamma _{\mu \nu }$ can be written as%
\begin{equation}
\gamma _{00}=-e^{f}=-\left( K-\frac{2\mu }{\varphi ^{(d-3)}}+\frac{\vartheta
^{2}}{\varphi ^{2(d-3)}}\right) ,  \tag{36}
\end{equation}%
where we used the definitions $K=\left( \frac{E}{A}+\frac{2DG}{A}+2\left( 
\frac{d-3}{d-2}\right) G^{2}\right) $, $\mu =\frac{D}{B(d-3)}+\frac{2CG}{%
(d-2)}$ and $\vartheta ^{2}=\frac{2C^{2}}{(d-2)(d-3)}$.

Since from (11) $\Omega =B^{-1}\dot{\varphi}\mathcal{F}^{-1}$ [see also eq.
(30)], we have that%
\begin{equation}
\gamma _{11}=e^{h}=B^{-1}\dot{\varphi}e^{-f}.  \tag{37}
\end{equation}

\smallskip

\noindent \textbf{4. Charged black holes and Hamiltonian formalism}

\smallskip

\noindent Now we turn our attention to a slightly different view of our
initial problem that resembles the analysis for a relativistic point
particle. This approach has been proved to be fruitful in several ways,
specially for quantization via the Dirac constrained systems formalism. The
Lagrangian (17) can be written in a more concise form%
\begin{equation}
\mathcal{L}=\frac{1}{2}\left \{ \lambda ^{-1}\dot{q}^{a}\dot{q}^{b}\xi
_{ab}+\lambda m_{o}^{2}\right \} .  \tag{38}
\end{equation}%
Here, we have defined the metric $\xi _{ab}$ with components%
\begin{equation}
\left[ \xi _{ab}\right] =%
\begin{pmatrix}
(d-3) & 1 & 0 \\ 
1 & 0 & 0 \\ 
0 & 0 & \frac{2}{(d-2)}e^{-2q^{2}}%
\end{pmatrix}%
.  \tag{39}
\end{equation}%
The inverse metric corresponding to (39) is%
\begin{equation}
\left[ \xi ^{ab}\right] =%
\begin{pmatrix}
0 & 1 & 0 \\ 
1 & -(d-3) & 0 \\ 
0 & 0 & \frac{1}{2}(d-2)e^{2q^{2}}%
\end{pmatrix}%
.  \tag{40}
\end{equation}%
Diagonalizing (39), we see that it has three eigenvalues, given by%
\begin{equation}
\begin{array}{c}
\phi _{1}=\frac{(d-3)+\sqrt{(d-3)^{2}+4}}{2}, \\ 
\\ 
\phi _{2}=\frac{(d-3)-\sqrt{(d-3)^{2}+4}}{2}, \\ 
\\ 
\phi _{3}=\frac{2}{(d-2)}e^{-2q^{2}}.%
\end{array}
\tag{41}
\end{equation}%
We note that in $d=4$, $\phi _{1}$ is precisely the golden ratio \cite%
{Nieto1}. Furthermore, in order to be consistent with the corresponding
eigenvectors, it is useful to define the new coordinates%
\begin{equation}
\begin{array}{c}
w^{1}=\frac{1}{\sqrt{\phi _{1}-\phi _{2}}}\left( \phi _{2}q^{1}+q^{2}\right)
, \\ 
\\ 
w^{2}=\frac{1}{\sqrt{\phi _{1}-\phi _{2}}}\left( \phi _{1}q^{1}+q^{2}\right)
, \\ 
\\ 
w^{3}=q^{3}.%
\end{array}
\tag{42}
\end{equation}%
In terms of these variables, (38) can be expressed as follows:%
\begin{equation}
\mathcal{L}=\frac{1}{2}\left \{ \lambda ^{-1}\dot{w}^{a}\dot{w}^{b}\eta
_{ab}+\lambda m_{o}^{2}\right \} .  \tag{43}
\end{equation}%
Although we have a similar form between (38) and (43), we have that now the
associated metric has the diagonal form $\eta _{ab}=diag\left( -1,1,\frac{2}{%
(d-2)}e^{-2q^{2}}\right) $.

Let us make some brief comments about the gauge fixing associated with (38).
As it has been just shown, the Lagrangian (38) has been expressed in a
similar form as the relativistic point particle, whose Lagrangian $L_{rel}$
goes as 
\begin{equation}
L_{rel}=\frac{1}{2}\Lambda ^{-1}\frac{dx^{\mu }}{d\tau }\frac{dx^{\nu }}{%
d\tau }\eta _{\mu \nu }-\frac{1}{2}\Lambda M_{0}^{2},  \tag{44}
\end{equation}%
where $\Lambda $ is a Lagrange multiplier, $M_{0}$ is the mass of the
system, $\eta _{\mu \nu }=diag(-1,1,...,1)$ and $\tau $ is an arbitrary
parameter. Varing (44) with respect to $\lambda $ leads to

\begin{equation}
\Lambda ^{-2}\frac{dx^{\mu }}{d\tau }\frac{dx_{\mu }}{d\tau }+M_{0}^{2}=0, 
\tag{45}
\end{equation}%
We observe that in this case (see Refs \cite{Govaerts} and \cite{Henneaux}
for details), since $M_{0}$ is a constant, it is useful to make the choice $%
\Lambda =\frac{1}{M_{0}}$. We get

\begin{equation}
\frac{dx^{\mu }}{d\tau }\frac{dx_{\mu }}{d\tau }=-1,  \tag{46}
\end{equation}%
or

\begin{equation}
dx^{\mu }dx_{\mu }=-d\tau ^{2}.  \tag{47}
\end{equation}%
In general we have the invariance of the line element

\begin{equation}
dx^{\mu }dx_{\mu }=dx^{^{\prime }\mu }dx_{\mu }^{\prime }.  \tag{48}
\end{equation}%
So, (47) can be obtained from (48) by setting $x^{^{\prime }0}=\tau $ and $%
x^{^{\prime }i}=0$. This prove that the choice $\Lambda =\frac{1}{M_{0}}$ is
equivalent to choose a reference system with $x^{^{\prime }i}=0$. It is
worth mentioning that this means that $\tau $ must be identified with the
proper time.

In our case, the expression (38) gives%
\begin{equation}
\lambda ^{-2}\dot{q}^{a}\dot{q}^{b}\xi _{ab}-m_{o}^{2}=0.  \tag{49}
\end{equation}%
Comparing (45) and (49) we observe that both equations are very similar.
However, while in (45) $M_{0}^{2}$ is a constant, one notes from (19) that $%
m_{0}^{2}$ is not. Moreover, while the $\eta _{\mu \nu }$ is diagonal $\xi
_{ab}$ is not. Hence we have more freedom to choose $\lambda $. A convenient
choice is

\begin{equation}
\lambda =\frac{(d-3)^{1/2}e^{-q^{1}}}{m_{0}}  \tag{50}
\end{equation}%
Thus, (49) leads to

\begin{equation}
dq^{a}dq^{b}\xi _{ab}=(d-3)e^{-2q^{1}}dr^{2}.  \tag{51}
\end{equation}%
Since one should have the invariant

\begin{equation}
dq^{a}dq^{b}\xi _{ab}=dq^{\prime a}dq^{\prime b}\xi _{ab}^{\prime }. 
\tag{52}
\end{equation}%
It is not difficult to see that (51) can be obtained from (52) by setting $%
dq^{\prime 2}=0$, $dq^{\prime 3}=0$ and taking $dq^{1}=dq^{\prime
1}=e^{-q^{1}}dr$. Since $\varphi =e^{q^{1}}$, this is equivalent to choose a
reference frame such that

\begin{equation}
d\varphi =dr.  \tag{53}
\end{equation}%
Let us set $\varphi =r$. With this choice it is straightforward to see from
(31) and (36) that both $dq^{\prime 2}=0$ and $dq^{\prime 3}=0$ can be
considered as conditions for $r\rightarrow \infty $. The condition $\varphi
=r$ is justified in the sense that it leads to the right Newtonian limit,
when one assumes asymptotic flatness. In this limit, in (36) one can set $%
K=1 $ and \cite{KunzNavarro}\cite{Farmany}%
\begin{equation}
\mu =\frac{4G_{d}\Gamma \left( \frac{d-1}{2}\right) }{(d-2)\pi ^{(d-3)/2}}M.
\tag{54}
\end{equation}%
Here $M$ is the mass of the black-hole and $\Gamma (n)$ is the gamma
function. Moreover, $\vartheta $ is related with the charge of the black
hole $Q$ (in Gaussian units) by mean of%
\begin{equation}
\vartheta ^{2}=\frac{2G_{d}}{(d-2)(d-3)}Q^{2}.  \tag{55}
\end{equation}%
To recover the usual Reissner-Nordstr\"{o}m solution, we set $B=1$ in (37).
Of course, in the limit where $Q=0$, the metric (2) corresponds to the usual
Schwarzschild-Tangherlini metric \cite{Tangherlini}.

By reviewing (36) and (37), we observe that two horizons arise, as is well
known, given by%
\begin{equation}
\varphi _{\pm }=\left( \mu \pm \sqrt{\mu ^{2}-\vartheta ^{2}}\right) ^{\frac{%
1}{(d-3)}}.  \tag{56}
\end{equation}

Until now, we have made the Lagrangian for a charged black hole in $d\geq 4$%
. We now turn our attention to the Hamiltonian analysis. The canonical
moments of $(q^{1},q^{2},q^{3})$ corresponding to (17) are given by $p_{a}=%
\frac{\partial \mathcal{L}}{\partial \dot{q}^{a}}$. Explicitly:

\begin{equation}
\begin{array}{c}
p_{1}=\frac{\partial \mathcal{L}}{\partial \dot{q}^{1}}=\lambda ^{-1}\left[
(d-3)\dot{q}^{1}+\dot{q}^{2}\right] , \\ 
\\ 
p_{2}=\frac{\partial \mathcal{L}}{\partial \dot{q}^{2}}=\lambda ^{-1}\dot{q}%
^{1}, \\ 
\\ 
p_{3}=\frac{\partial \mathcal{L}}{\partial \dot{q}^{3}}=\frac{2}{(d-2)}%
\lambda ^{-1}e^{-2q^{2}}\dot{q}^{3}.%
\end{array}
\tag{57}
\end{equation}%
By defining the Legendre transform of (38) as%
\begin{equation}
\mathcal{H}_{0}\mathcal{=}\dot{q}^{a}p_{a}-\mathcal{L},  \tag{58}
\end{equation}%
we obtain from (57) that%
\begin{equation}
\mathcal{H}_{0}=\frac{\lambda }{2}\left \{ \left[ 2p_{1}p_{2}-(d-3)\left(
p_{2}\right) ^{2}+\frac{1}{2}(d-2)e^{2q^{2}}\left( p_{3}\right) ^{2}\right]
-m_{o}^{2}\right \} .  \tag{59}
\end{equation}%
This can be shown to be equivalent to%
\begin{equation}
\mathcal{H}_{L}=\mathcal{L-}\lambda m_{o}^{2}.  \tag{60}
\end{equation}%
Also, the Lagrangian (38) can be written as 
\begin{equation}
\mathcal{L}^{\prime }=\dot{q}^{a}p_{a}-\frac{\lambda }{2}\left \{ \left[
2p_{1}p_{2}-(d-3)\left( p_{2}\right) ^{2}+\frac{1}{2}(d-2)e^{2q^{2}}\left(
p_{3}\right) ^{2}\right] -m_{o}^{2}\right \} ,  \tag{61}
\end{equation}%
or using (40), as%
\begin{equation}
\mathcal{L}^{\prime }=\dot{q}^{a}p_{a}-\frac{\lambda }{2}\left \{
p_{a}p_{b}\xi ^{ab}-m_{o}^{2}\right \} .  \tag{62}
\end{equation}

It is suggestive to call the expression inside the bracket $\mathcal{H}_{L}$%
, since the variation of (62) respect to $\lambda $ yields%
\begin{equation}
\mathcal{H}_{L}=p_{a}p_{b}\xi ^{ab}-m_{o}^{2}=0.  \tag{63}
\end{equation}%
This constraint can be obtained by substituting (57) in $\mathcal{H}_{L}$,
that yields the same relation (23).

From here, it would interesting to explore the implications of applying the
Dirac's quantization formalism for constrained systems.\bigskip

\smallskip

\noindent \textbf{5. Final remarks}

\smallskip

\noindent In this work we have obtained the known Reisner-Nordstr\"{o}m
black hole solution for arbitrary dimensions ($d\geq 4$). This solution was
obtained by mean of a Lagrangian approach that had not been explored in the
literature until very recently \cite{Nieto1}.

More precisely, starting with the most general metric associated to radial
symmetry, we proceeded to consider the problem as an analogue to that of a
relativistic point particle. For this purpose, a convenient choice of
independent coordinates allowed us to take the analysis with a reduced
metric. We argue that the transformations performed\ makes the new metric
suitable for quantization according to the Dirac's constrained systems
formalism. Although this idea is well known and form the basis for several
approaches in quantum gravity [23]-[25], our method for obtaining the
solutions, as well as the implications that it can have for quantization, do
not seem to have been explored before.

We showed that our procedure applies for both Reisner-Nordstr\"{o}m and
Schwarzschild solutions in $d>3$. The applicability of the method appears to
be directly related to spherical symmetry, that in turns leads to treat the
static models (i.e. Schwarzschild-Tangherlini \ and Reissner-Nordstr\"{o}m)
as constrained systems with only one first class constraint. Nevertheless,
it is interesting to explore also the case of stationary solutions from this
perspective. In this case, in order to obtain the Kerr-Newman or Myers-Perry
metrics with our procedure, we expect that the introduction of new
constraints permits to get rid of effects like frame dragging, simplifying
thus the analysis \cite{Myers}\cite{Gibbons}.

Also, as we have seen, the initial problem was simplified to a reduced
metric that can be diagonalized, in such a way that it can be suitable for
analysing quantum aspects of black holes \cite{Strominger}\cite{Vercnocke}.
Both the procedure and the analysis used in this work can have implications
to both cosmological models in higher dimensions and alternative theories of
gravity (see for instance [26]-[30] as well as references therein). This
type of explorations will be reported elsewhere.\smallskip

\noindent \textbf{Acknowledgments: }JA and EAL recognize that this work was
partially supported by PROFAPI-UAS, 2011. VMV would like to thank the
Departamento de F\'{\i}sica, Universidad de Guanajuato, for hospitality
during a stage of this work, and also to the Coordinaci\'{o}n de Investigaci%
\'{o}n (CIC) de la UMSNH for support.

\noindent \textbf{Appendix. The Riemann and Ricci tensor.}\smallskip \ 

\noindent From the metric (2), we find that the only nonvanishing
Christoffel symbols, defined by $\Gamma _{\alpha \beta }^{\sigma }=\frac{1}{2%
}g^{\sigma \lambda }(g_{\lambda \beta ,\alpha }+g_{\alpha \lambda ,\beta
}-g_{\alpha \beta ,\lambda })$, are%
\begin{equation}
\begin{array}{ccccc}
\Gamma _{01}^{0}=\frac{1}{2}\dot{f} &  & \Gamma _{00}^{1}=\frac{1}{2}\dot{f}%
e^{f-h} &  & \Gamma _{11}^{1}=\frac{1}{2}\dot{h} \\ 
&  &  &  &  \\ 
\Gamma _{ij}^{1}=-e^{-h}\varphi \dot{\varphi}\tilde{\gamma}_{ij} &  & \Gamma
_{1j}^{i}=\varphi ^{-1}\dot{\varphi}\delta _{j}^{i} &  & \Gamma _{jk}^{i}=%
\tilde{\Gamma}_{jk}^{i}.%
\end{array}
\tag{A.1}
\end{equation}%
The Riemann tensor is defined by $R_{~\mu \beta \nu }^{\alpha }=\partial
_{\beta }\Gamma _{\mu \nu }^{\alpha }-\partial _{\nu }\Gamma _{\mu \beta
}^{\alpha }+\Gamma _{\beta \lambda }^{\alpha }\Gamma _{\mu \nu }^{\lambda
}-\Gamma _{\nu \lambda }^{\alpha }\Gamma _{\mu \beta }^{\lambda }$, and we
have as relevant components, according to (A.1): 
\begin{equation}
\begin{array}{ccc}
R_{~101}^{0}=-\frac{1}{2}\ddot{f}+\frac{1}{4}\dot{f}\dot{h}-\frac{1}{4}\dot{f%
}^{2}, &  & R_{~010}^{1}=-e^{f-h}\left( -\frac{1}{2}\ddot{f}+\frac{1}{4}\dot{%
f}\dot{h}-\frac{1}{4}\dot{f}^{2}\right) , \\ 
&  &  \\ 
R_{~i0j}^{0}=-\frac{1}{2}e^{-h}\tilde{\gamma}_{ij}\varphi \dot{\varphi}\dot{f%
}, &  & R_{~0j0}^{i}=\frac{1}{2}e^{f-h}\delta _{j}^{i}\varphi ^{-1}\dot{%
\varphi}\dot{f}, \\ 
&  &  \\ 
R_{~i1j}^{1}=e^{-h}\tilde{\gamma}_{ij}\left( \frac{1}{2}\varphi \dot{\varphi}%
\dot{h}-\varphi \ddot{\varphi}\right) , &  & R_{~1j1}^{i}=\delta
_{j}^{i}\varphi ^{-1}\left( \frac{1}{2}\dot{\varphi}\dot{h}-\ddot{\varphi}%
\right) , \\ 
&  &  \\ 
R_{~jkl}^{i}=\tilde{R}_{~jkl}^{i}+e^{-h}\dot{\varphi}^{2}\left( \delta
_{l}^{i}\tilde{\gamma}_{jk}-\delta _{k}^{i}\tilde{\gamma}_{jl}\right) . &  & 
\end{array}
\tag{A.2}
\end{equation}

Next, the Ricci tensor $R_{\mu \nu }=R_{~\mu \alpha \nu }^{\alpha }$ has
components%
\begin{equation}
\begin{array}{c}
R_{00}=e^{f-h}\left( \frac{1}{2}\ddot{f}-\frac{1}{4}\dot{f}\dot{h}+\frac{1}{4%
}\dot{f}^{2}+\frac{D-2}{2}\varphi ^{-1}\dot{\varphi}\dot{f}\right) , \\ 
\\ 
R_{11}=-\frac{1}{2}\ddot{f}+\frac{1}{4}\dot{f}\dot{h}-\frac{1}{4}\dot{f}%
^{2}+(D-2)\varphi ^{-1}\left( \frac{1}{2}\dot{\varphi}\dot{h}-\ddot{\varphi}%
\right) , \\ 
\\ 
R_{ij}=e^{-h}\tilde{\gamma}_{ij}\left[ -\frac{1}{2}\varphi \dot{\varphi}\dot{%
f}+\frac{1}{2}\dot{h}\varphi \dot{\varphi}-\varphi \ddot{\varphi}-(d-3)\dot{%
\varphi}^{2}+k(d-3)e^{h}\right] .%
\end{array}
\tag{A.3}
\end{equation}%
We note that we have used the fact that the maximally symmetric subspace
defined by the metric $\tilde{\gamma}_{ij}$ demands that $\tilde{R}_{ij}=%
\tilde{\gamma}^{kl}\tilde{R}_{kilj}=k(d-3)\tilde{\gamma}_{ij}$. This in turn
implies that $\tilde{R}=k(d-2)(d-3)$, that is used to obtain the curvature
tensor $R=g^{\mu \nu }R_{\mu \nu }$ as%
\begin{equation}
\begin{array}{c}
R=e^{-h}(d-2)\left[ \varphi ^{-1}\dot{\varphi}(\dot{h}-\dot{f})-2\varphi
^{-1}\ddot{\varphi}-(d-3)\varphi ^{-2}\dot{\varphi}^{2}\right] + \\ 
\\ 
+e^{-h}\left( -\ddot{f}+\frac{1}{2}\dot{f}\dot{h}-\frac{1}{2}\dot{f}%
^{2}\right) +k(d-2)(d-3)\varphi ^{-2}.%
\end{array}
\tag{A.4}
\end{equation}

\end{document}